\documentclass[a4paper, twoside]{article}

\usepackage[T1]{fontenc}
\usepackage[utf8]{inputenc}

\usepackage{graphicx}
\usepackage{subfig}
\usepackage[export]{adjustbox}
\usepackage{wrapfig}
\usepackage{tikz}
\usepackage{pgfplots}
\pgfplotsset{compat=1.18}

\definecolor{bloqadegreen}{RGB}{27,158,119} 
\usepackage{amsmath,amssymb,amsfonts}
\usepackage{mathtools}
    \DeclarePairedDelimiter\bra{\langle}{\rvert}
    \DeclarePairedDelimiter\ket{\lvert}{\rangle}
    \DeclarePairedDelimiterX\braket[2]{\langle}{\rangle}{#1\,\delimsize\vert\,\mathopen{}#2}
\usepackage{amsthm}
    \theoremstyle{definition}
    \newtheorem{definition}{Definition}[section]
    \theoremstyle{definition}

\usepackage[margin=0.8in]{geometry}
\usepackage{algorithmic}
\usepackage{textcomp}
\usepackage{float}
\usepackage{cancel}
\usepackage{fancyhdr}
\usepackage[hang,flushmargin]{footmisc}
\usepackage{multicol}
\usepackage{abstract}

\usepackage{layout}

\usepackage{program}
\usepackage[]{xcolor}
\usepackage{lettrine}
\usepackage{cite}

\usepackage{siunitx}
\usepackage{titlesec}
    \titleformat*{\section}{\large\scshape\bfseries}
    \titleformat*{\subsection}{\large\slshape\rmfamily}
    \titleformat*{\subsubsection}{\large\scshape\sffamily}
    \titleformat*{\paragraph}{\large\scshape\sffamily}
    \titleformat*{\subparagraph}{\large\scshape\sffamily}

\makeatletter
\newcommand{\varnotsuccsim}{\mathrel{\mathpalette\varn@t\succsim}}
\newcommand{\varn@t}[2]{%
  \vphantom{/{#2}}%
  \ooalign{\hfil$\m@th#1/\mkern2mu$\cr\hfil$\m@th#1#2$\hfil\cr}%
}
\makeatother

\def\BibTeX{{\rm B\kern-.05em{\sc i\kern-.025em b}\kern-.08em
    T\kern-.1667em\lower.7ex\hbox{E}\kern-.125emX}}

\makeatletter
\def\@fnsymbol#1{\ensuremath{\ifcase#1\or \dagger\or \ddagger\or
    \mathsection\or *\or \mathparagraph\or \|\or **\or \dagger\dagger
    \or \ddagger\ddagger \else\@ctrerr\fi}}
\makeatother

\usepackage[unicode=true,pdfusetitle,
 bookmarks=true,bookmarksnumbered=false,bookmarksopen=false,
 breaklinks=false,pdfborder={0 0 0},pdfborderstyle={},backref=false,colorlinks=true, 
 linkcolor=blue, urlcolor=magenta, citecolor=green, pdfstartview={FitH}, 
 hyperfootnotes=false]{hyperref}
\date{}
\begin{document}

\title{Mapping Game Theory to Quantum Systems: Nash Equilibria via Neutral Atom Computing}

\author{
    D. Di Gregorio\thanks{Master's student in Physics of Complex Systems at Polytechnic University of Turin.},
    G. Ferrannini\footnotemark[1],
    F. Fissore\thanks{Master's student in Quantum Engineering at Polytechnic University of Turin.}.
}

\onecolumn
\maketitle
\begin{center}
    \textit{Revised from the IEEE R8 Student Paper Contest 2025 submission}
\end{center}

\begin{abstract}
    Nash equilibria are crucial for understanding game behavior and systems 
    in economics, physics, biology, and computer science. A significant 
    application arises from the connection between Nash equilibria and 
    optimization problems . 
    However, finding Nash equilibria is challenging due to its NP-Hard 
    complexity, specifically within the PPAD class. 
    By exploiting the correspondence between Maximum Independent Sets (MIS) and Nash equilibria on unit-disk graphs, we map these problems onto the ground state configurations of Rydberg atom arrays. Simulations show the effectiveness of this quantum method, highlighting its potential for solving complex problems in game theory.
\end{abstract}

\begin{multicols}{2}
\section{Introduction}
\label{sec:introduction}
\lettrine[findent=2pt]{\textbf{I}}{ } n recent decades, the increasing complexity of models in economics, physics, and natural sciences -- such as biology and chemistry -- has necessitated the development of advanced algorithms and mathematical theories to solve complex games on networks. Despite the strict assumption of rational players, game theory has revolutionized our understanding of strategic interactions and decision-making for nearly a century. A pivotal advancement in this field was the introduction of the concept of \textit{mixed equilibrium}, commonly known as the \textit{Nash equilibrium}, first formalized by Nash \cite{Nash1951NONCOOPERATIVEG}. Since its inception, significant efforts have been devoted to devising methods for evaluating the set of equilibria in a game. However, the complexity of finding the set of Nash equilibria is so high that it often renders the task intractable for very large games. The Nash-equilibrium-finding problem belongs to a subclass of the NP-hard complexity class. While it is not an NP-complete problem due to Nash's proof of guaranteed solutions in convex games \cite{Nash1951NONCOOPERATIVEG}, it is generally classified as a PPAD problem \cite{GILBOA198980,Complexity_of_Nash_proceedings,Complexity_of_Nash_article,Rcz2011LectureC}. To address these challenges, many approximate methods \cite{Rcz2011LectureC} and algorithms \cite{Facchinei_Pang_2009,MCKELVEY_algo_for_Nash,Parsopoulos_algo_Nash,Herings-Peeters_algo_for_Nash} have been developed. Another promising direction involves exploiting the topological features of networks, which are closely related to the properties of the games performed on them. Notably, there exists a strong connection between \textit{maximum independent sets} (MIS) in graphs and Nash equilibria \cite{DallAsta_Nash_and_MIS,BRAMOULLE2007478}. The problem of finding maximal independent sets (mIS) in graphs has been extensively studied and is known to be NP-hard \cite{Lozin2007MaximumIS,Alekseev_MIS}. The advent of quantum computation has opened new horizons in tackling these computational challenges. Specifically, it has been demonstrated that it is possible to map a coordination game on a planar graph onto a neutral-atom quantum computer. By retrieving the MIS through determining the system’s ground state, the Nash equilibrium can be effectively computed \cite{Dalyac2024_MIS_with_NA}. This process involves minimizing the Hamiltonian of the quantum system, showcasing a novel and promising approach to solving traditionally intractable problems in game theory.
\paragraph*{This work and main findings}  
We leverage the MIS–Nash connection on unit-disk graphs to map public-good games on networks to Rydberg-atom arrays. Concretely:  
(i) we formalize the correspondence between specialized equilibria and (maximal) independent sets and adopt it as the bridge to quantum hardware;  
(ii) we instantiate the mapping on a neutral-atom platform by specifying the geometric embedding and annealing schedule;  
(iii) we perform Bloqade simulations on two representative instances: Graph~A (unique MIS) and Graph~B (four MIS); both shown in Figure~\ref{fig:1}.  
Across 1000-shot runs, the most frequent outcomes concentrate on MIS configurations. For Graph~A (Figure ~\ref{fig:subimage3}), the dominant readout coincides with the unique MIS; for Graph~B (Figures ~\ref{fig:subimage4}-\ref{fig:subimage7}), the four MIS appear among the highest-count configurations, with deviations attributable to diabatic transitions and noise.  
(iv) We classically verify all Nash equilibria and mIS, observing a one-to-one agreement with the quantum outputs for the targeted configurations.  
(v) We discuss mapping constraints (geometry, hardware spacing, and blockade radius), as well as computational-time considerations arising from scheduling and decoherence.
\paragraph*{Outline}  
Section~\ref{sec: Preliminaries} recalls basic game-theoretic notions and the public-goods-on-networks model, consolidating definitions used later. Section~\ref{sec: Basic} introduces the Hamiltonian, the Rydberg blockade, and the MIS/ground-state correspondence relevant to unit-disk embeddings. Section~\ref{sec:Analysis} presents the analysis pipeline and simulations: we detail the embedding, annealing profiles, and shot statistics, and we compare quantum outputs with classical Nash and mIS enumerations. The subsection~\ref{subsec:Map} synthesizes technological and graph-theoretic constraints for feasible embeddings. We then discuss computational-time aspects and their dependence on hardware and spectral gaps. Finally, Section~\ref{sec:conclusions} concludes with implications and perspectives for quantum approaches to equilibrium computation.

\begin{figure}[H]
  \centering
  \includegraphics[width=0.5\textwidth]{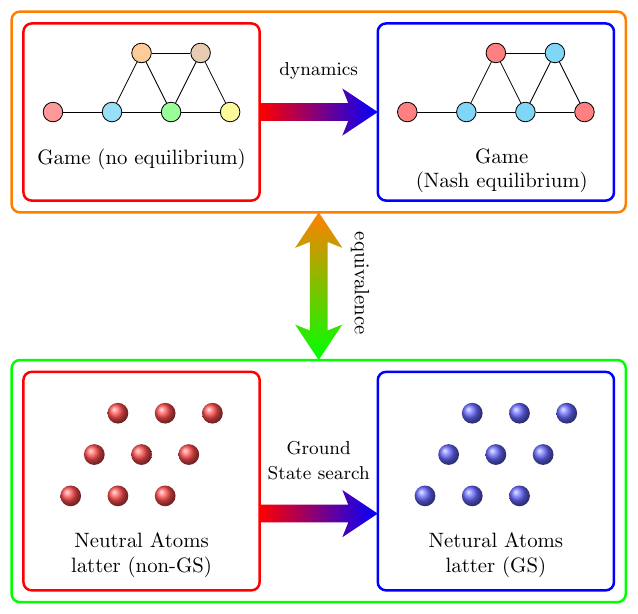}
  \caption*{Pictorial abstract}
\end{figure}

\section{Preliminaries}
\label{sec: Preliminaries}
Before diving into the application of quantum computation to Nash equilibria 
and analyzing the main results, it is important to present some simple 
definitions in Game Theory.

\subsection{Basics definitions in Game Theory}
A \textit{game} is a formalized interaction between rational agents (players) making 
decisions, often in the presence of competition or cooperation. Game theory is the 
multi-agent extension of decision theory, explaining how self-interested, rational 
agents make decisions based on others' behaviors. To understand it, we start by 
considering a set of mutually exclusive \textit{alternatives}, where each player 
ranks or chooses among options. Rational players aim to maximize their utility functions
(also called payoffs), 
influenced by others' actions.

\begin{definition}\label{def:cardinal}
    A \textit{cardinal} (or \textit{interval-valued}) \textit{utility function} 
    that represents the preference relation $\succsim$ is unique up to a transformation 
    such that if $u$ is a cardinal utility function for $\succsim$, then 
    $\tilde{u} = a + b u$, with $a, b \in \mathbb{R}$ and $b > 0$, is also a cardinal utility 
    function.
\end{definition}
\noindent The term \textit{cardinal} differentiates this utility concept from 
\textit{ordinal} utility, which is not relevant here. In a game, each player has a 
utility function dependent on their own and others' actions. When multiple agents 
make decisions affecting each other’s utility in order to maximise their own, the game is called 
\textit{non-cooperative}. A \textit{best response} is defined as the optimal action 
for a player, given the others' strategies.

\begin{definition}\label{def:Best_Response}
    The \textit{pure-strategy best response} of player $i$ to the action profile 
    $a_{-i}$ is an action $a^{*}_{i} \in A_i$ such that 
    $u_i(a^*_i, a_{-i}) \geq u_i(a_i, a_{-i}), \forall a_i \in A_i$,
    where $A_i$ is the set of actions available.\\
    Thus, we can define the pure-strategy best response as
    \begin{align*}
        \mathcal{B}_i&(a_{-i})=\\
        &=\{a_i\in A_i : u_i(a_i,a_{-i})\geq u_i(a'_i,a_{-i}),\forall a'_i \in A_i\}
    \end{align*}
\end{definition}
\noindent If all players play the best response to the actions played by the others, 
it may be possible to find an equilibrium point, called \textit{Nash equilibrium}.
\begin{definition}
    A \textit{pure-strategy Nash equilibirum} for a game $G = (\mathcal{N},A,u)$ 
    is an \textit{action profile}, i.e., a \textit{strategy}, $\bar{a}^*=(a^*_1,a^*_2,\dots)$ such that 
    $$a^*_i\in\mathcal{B}_i(a^*_{-i}),\forall i\in\mathcal{N}$$
    or, in other terms
    $$\forall A_i\in\mathcal{N},\forall a'_i \in A_i, u_i(a^*_i,a^*_{-i})\geq u_i(a'_i,a^*_{-i})$$
\end{definition}
\noindent Meaning that \textit{in a Nash equilibrium, no player finds it profitable to 
unilaterally deviate from the equilibrium strategy profile}.
If a player deviates, the equilibrium may be disrupted or a new one may emerge. These 
games involve \textit{pure strategies}, where players choose actions with certainty, 
described in \textit{normal} or \textit{strategic} form.

\begin{definition}
    A finite $N$-person \textit{normal-form game} is a tuple $(\mathcal{N}, A, u)$ 
    in which
    \begin{itemize}
        \item $\mathcal{N}$ is a \textit{finite} set of $N$ players;
        \item $A = \times_{i=1,\dots,N} A_i$, where $A_i$ is a \textit{finite} 
        set of actions available to player $i$;
        \item $u = (u_1, \dots, u_N)$, where $u_i: A \to \mathbb{R}$ is a real-valued utility 
        function for player $i$.
    \end{itemize}
\end{definition}

\subsection{Public Goods in Networks Game}
In their article \cite{BRAMOULLE2007478}, Bramoullé and Kranton present a model of public goods in networks where agents decide whether to contribute to a public good that benefits themselves and their direct neighbors. The good is non-excludable, meaning that once it is produced, others can benefit from it without restriction. This leads effort levels to be viewed as strategic substitutes: an agent’s contribution depends on the contributions of its direct neighbors. The more neighbors contribute, the less incentive an agent has to do the same, as it can free-ride on their efforts. In \cite{BRAMOULLE2007478}, a particular type of profile is of critical interest and is defined as follows:

\begin{definition}
    A profile (strategy) $e$ is called \textit{specialized} if every agent either contributes the maximum effort level $e^*$ or does not contribute at all $(e = 0)$. Formally, for every agent $i$, $e_i = 0$ or $e_i = e^*$ \cite{BRAMOULLE2007478}.
\end{definition}

\noindent As shown in Theorem 1 of \cite{BRAMOULLE2007478}, a specialized profile is a Nash equilibrium if and only if its set of specialists (contributing agents) forms a maximal independent set.

\begin{definition}
    An independent set $I$ in a graph $g$ is a subset of nodes such that no two nodes in $I$ are connected. An independent set is \textit{maximal} if it is not a proper subset of any other independent set.
\end{definition}

\noindent This definition has significant implications for the game, as it establishes the following:
\begin{itemize}
    \item Nodes that contribute cannot be connected to each other.
    \item Every node that does not contribute must be connected to at least one contributing node. Thus, non-contributing nodes have no incentive to change their strategy, as they already benefit from their neighbors’ contributions.
\end{itemize}
This perfectly aligns with the property of strategic substitutability, where agents who specialize cannot be linked to each other in equilibrium. Consequently, this leads to the formation of an independent set and establishes a one-to-one correspondence between maximal independent sets (mIS) and Nash equilibria in this type of game \cite{DallAsta_Nash_and_MIS, BRAMOULLE2007478}.

\begin{definition}
    Among all maximal independent sets (mISs), a \textit{Maximum Independent Set} (MIS) is an mIS that maximizes the cardinality of the independent set. In other words, an MIS is an independent set in a graph that contains the largest possible number of nodes compared to all other independent sets.
\end{definition}

\noindent In the context of public goods, MISs represent Nash equilibria where the number of contributors is maximized. These equilibria are particularly noteworthy as they minimize free-riding compared to other equilibria.

\subsection{Pay-off in the model}
The payoff function for a single agent i is defined as follows

\begin{equation*}
    U_{i}(e;g)=b\Bigg(e_i + \sum_{j\in N_i}e_j\Bigg) - c\cdot e_i
\end{equation*}
where
\begin{itemize}
    \item $e=(e_1,e_2,\dots,e_n)$ represents the effort profile of all agents 
        in the network;
    \item $e_i$ is the effort chosen by agent $i$;
    \item $N_i$ is the direct neighborhood of $i$ in the network;
    \item $\sum_{j\in N_i}e_j$ is the total effort contributed by agent 
        $i$’s neighbors;
    \item $b(\cdot)$ is a concave benefit function derived from the 
        public good;
    \item $c\cdot e_i$ represents the linear cost of effort for agent $i$, 
    where $c > 0$ is the marginal cost.
\end{itemize}

This function balances the benefits $b$ from access to the public good (produced by 
the agent’s and their neighbors’ efforts) with the costs $c \cdot e_i$ of individual 
effort \cite{BRAMOULLE2007478}. An agent’s payoff in a networked public goods game 
depends on the benefit from the public good, the cost of contribution, and the 
locality of benefits. Contributing is advantageous only when neighbors’ 
contributions are insufficient. A Nash equilibrium occurs when contributing nodes are 
positioned so that no agent has any incentive to improve its payoff by increasing 
contributions, and non-contributing nodes benefit from others’ efforts without 
incurring costs.

\section{Basics of Neutral atom quantum computing}
\label{sec: Basic}
\subsection{Hamiltonian and parameters}
In this context, the Hamiltonian is delineated as a cost function that must be optimized.
For a more detailed discussion on this topic, see \cite{NA_quantum}.
Notice that the Hamiltonian in the cited paper is a general expression for neutral atom 
quantum computers and, as such, is not tailored specifically to the details of 
the system discussed here. 
It encapsulates the broad interactions typically encountered in such 
systems, but it should be understood as a flexible model rather than a precise 
description of the specific parameters and interactions of our setup.
Indeed, by defining $\ket{0}$ as the ground state and $\ket{1}$ as the Rydberg state, 
the resulting quantum dynamics are governed by the Hamiltonian

\begin{align*}
    \mathcal{H}(t) &= \frac{\hbar}{2}\Omega(t)\sum_i\big[\ket{0}_i\bra{1}_i + \text{h.c.}\big]+ \\
    &-\hbar\Delta(t)\sum_i n_i + \sum_{i<j}\mathcal{V}_{ij}n_i n_j
\end{align*}

\noindent where $n_i = \ket{\cdot}_i\bra{\cdot}_i$ is the quantum number operator, 
$\Delta$ is the detuning parameter, $\Omega$ is the Rabi frequency, and $\mathcal{V}_{ij}$  is the interaction strength \cite{Aquila_Quera,NA_quantum}. We will assume homogeneous coherent coupling, defined by a uniform Rabi frequency 
$(\lvert\Omega_i\rvert = \Omega)$ and detuning $(\Delta_i = \Delta)$. 
These parameters are dynamically controlled through adjustments in the intensity and detuning of the driving lasers. To prepare the system, the detuning $\Delta(t)$ is adjusted adiabatically over time while $\Omega$ is turned-on, transforming the initial ground state  
into the target one \cite{Bernien2017}.

\subsection{Rydberg Blockade}
When neutral atoms are coherently coupled to highly excited Rydberg states, they experience repulsive van der Waals interactions. 
The interaction potential between a pair of Rydberg atoms is expressed as:
\[
V_{ij} = \frac{C}{R_{ij}^6},
\]
where $R_{ij}$ represents the distance between the two atoms, and $C > 0$ is the van der Waals coefficient \cite{Jaksch2000}. The strength of the interaction, $V_{ij}$, can be adjusted by varying the distance between the atoms or by coupling them to different Rydberg states. 
These strong, coherent interactions impose a restriction known as the Rydberg blockade  \cite{Jaksch2000}, which prevents simultaneous excitation of nearby atoms into 
Rydberg states. 
This phenomenon occurs when $V_{ij}$ exceeds the effective Rabi frequency $\Omega$, effectively suppressing multiple Rydberg excitations. 
The spatial region within which this constraint applies is characterized by the Rydberg blockade radius, $R_b$. If both detuning and Rabi drive are nonzero, then the two effects can be combined as a characteristic 
energy scale to define the blockade radius as: 

\begin{equation*}
   R_b= \left(\frac{C}{\sqrt{\Omega^2+\Delta^2}}\right)^\frac{1}{6}
\end{equation*}

\subsection{MIS and Ground State correspondence}
One of the most significant subclasses of the Maximum Independent Set (MIS) problem 
is its formulation on a unit-disk graph. A unit-disk graph is defined as a geometric 
graph where vertices correspond to points in the plane, and edges connect any two vertices 
separated by a Euclidean distance less than or equal to a fixed radius. Computing the MIS 
on unit-disk graphs is NP-hard in the worst case, highlighting the computational challenges 
associated with solving this problem \cite{CLARK1991165}.
Neutral-atom quantum computing provides a powerful framework for addressing the MIS problem 
on unit-disk graphs by leveraging the connection between the unit-disk radius $R_{ud}$ 
and the Rydberg blockade radius $R_b$, where $R_b = R_{ud}$. In this setup, the ground state 
of the Hamiltonian encodes the MIS, as the atoms excited to the Rydberg state represent the 
independent set \cite{Aquila_Quera}. With a positive detuning $\Delta$, adding Rydberg excitations lowers the ground state energy. 

However, the Rydberg excitations are subject to the Rydberg blockade effect --  a large positive energy penalty 
that enforces that only one Rydberg excitation is allowed within the blockade radius\cite{Aquila_Quera}. The interplay of these 
two mechanisms is at the core of this correspondence, effectively preventing excited atoms from having other excited atoms as neighbors (forming an independent set). 
The graph structure is directly mapped onto the hardware by positioning neutral atoms at scaled coordinates that match the vertices of the target graph and setting the desired blockade radius $R_b$ through an appropriate choice of the parameters $\Omega$ and $\Delta$, as illustrated by the unit-disk embeddings shown in Fig.~\ref{fig:unitdisk_AB}.
This approach effectively allows the system to find a MIS through its adiabatic 
evolution \cite{Rydberg_for_MIS}.
\vspace*{-0.4\baselineskip}
{
\setlength{\intextsep}{-10pt}      
\setlength{\textfloatsep}{4pt}    
\begin{figure}[H]
  \centering
  \begingroup
    \setlength{\abovecaptionskip}{3pt}%
    \setlength{\belowcaptionskip}{0pt}%

    \subfloat[Graph A]{%
      \includegraphics[width=0.7\columnwidth]{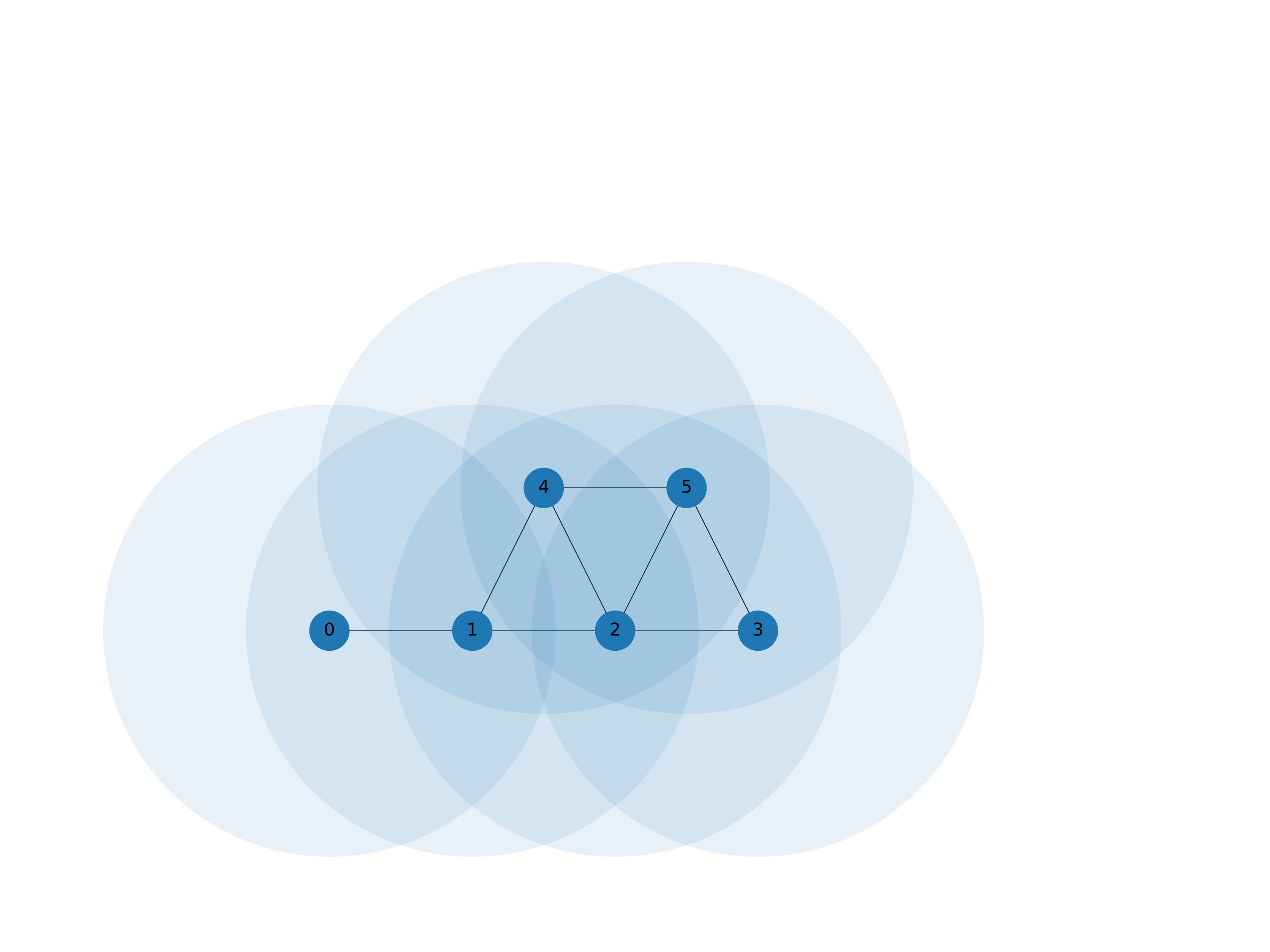}
    }\\[-0.6ex] 

    \subfloat[Graph B]{%
      \includegraphics[width=0.55\columnwidth]{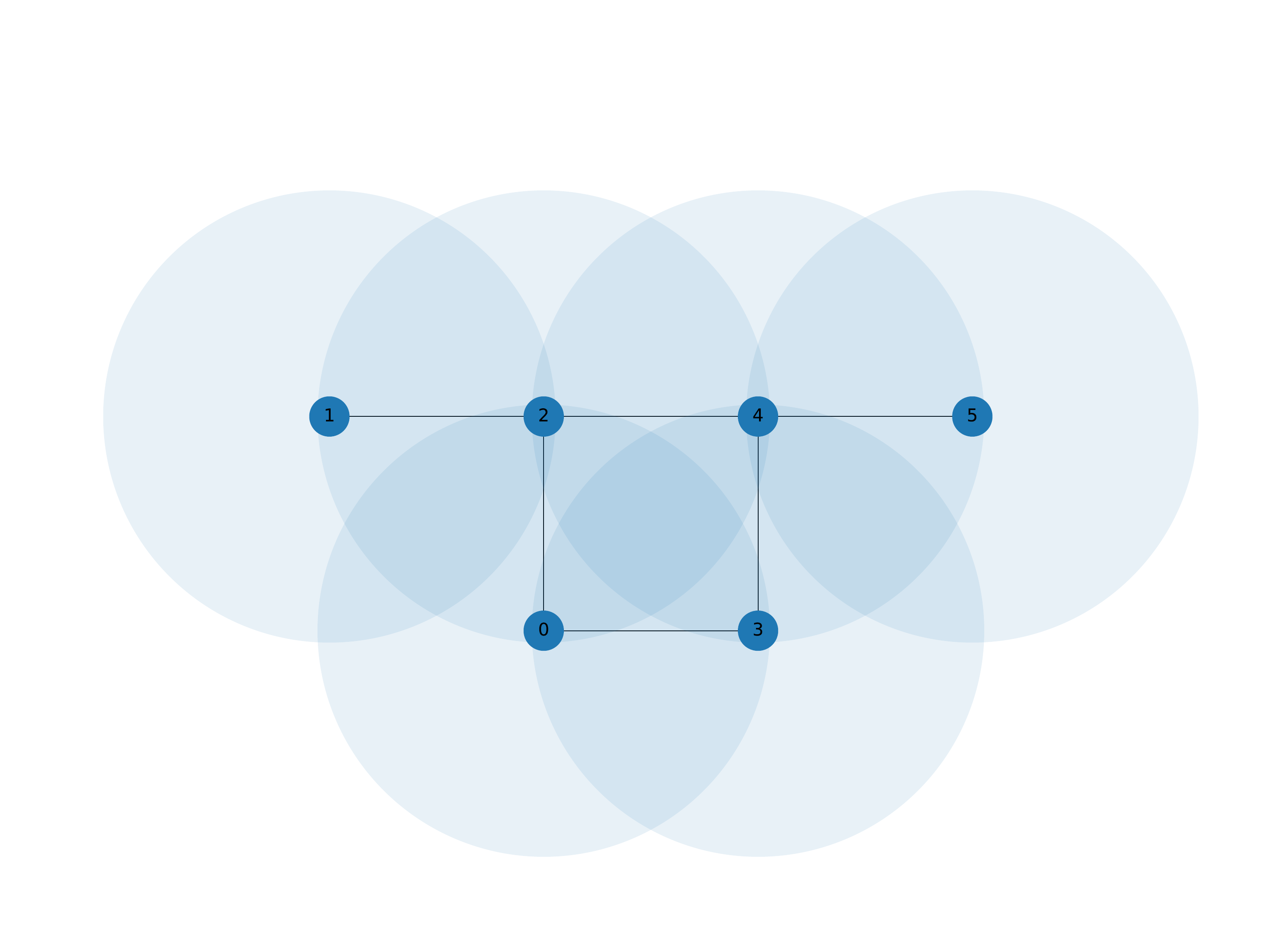}
    }

    \caption{Unit-disk graph embeddings for the two instances considered in the simulations. 
    The disks of radius $R_b$ enforce the independence constraint corresponding to the Rydberg blockade.}
    \label{fig:unitdisk_AB}
  \endgroup
\end{figure}
}

\begin{figure}[H]
\centering

\subfloat[Graph A\label{fig:subimage1}]{%
\begin{tikzpicture}[
  scale=0.11, 
  edge/.style={gray, line width=0.7pt},
  vertex/.style={circle, inner sep=1.8pt},
]

\newcommand{\graphA}[6]{%
  \coordinate (a1) at (0,0);
  \coordinate (a2) at (6,0);
  \coordinate (a3) at (12,0);
  \coordinate (a4) at (18,0);
  \coordinate (a5) at (9,{3*sqrt(3)});
  \coordinate (a6) at (15,{3*sqrt(3)});

  \draw[edge] (a1)--(a2)--(a3)--(a4);
  \draw[edge] (a2)--(a5)--(a3);
  \draw[edge] (a3)--(a6)--(a4);
  \draw[edge] (a5)--(a6);

  \node[vertex, fill=#1] at (a1) {};
  \node[vertex, fill=#2] at (a2) {};
  \node[vertex, fill=#3] at (a3) {};
  \node[vertex, fill=#4] at (a4) {};
  \node[vertex, fill=#5] at (a5) {};
  \node[vertex, fill=#6] at (a6) {};
}

\begin{scope}
  \begin{scope}[shift={(0,0)}]
    \graphA{red}{blue}{blue}{red}{red}{blue}
  \end{scope}

  \begin{scope}[shift={(24,0)}]
    \graphA{blue}{red}{blue}{red}{blue}{blue}
  \end{scope}

  \begin{scope}[shift={(48,0)}]
    \graphA{red}{blue}{blue}{blue}{blue}{red}
  \end{scope}

  \begin{scope}[shift={(12,-16)}]
    \graphA{blue}{red}{blue}{blue}{blue}{red}
  \end{scope}

  \begin{scope}[shift={(36,-16)}]
    \graphA{red}{blue}{red}{blue}{blue}{blue}
  \end{scope}

\end{scope}

\end{tikzpicture}
}\\[1.5ex]

\subfloat[Graph B\label{fig:subimage2}]{%
\begin{tikzpicture}[
  scale=0.55,
  edge/.style={gray, line width=0.7pt},
  vertex/.style={circle, inner sep=1.8pt},
]

\newcommand{\graphB}[6]{%
  \coordinate (t1) at (0,0);
  \coordinate (t2) at (1,0);
  \coordinate (t3) at (2,0);
  \coordinate (t4) at (3,0);
  \coordinate (b1) at (1,-1);
  \coordinate (b2) at (2,-1);

  \draw[edge] (t1)--(t2)--(t3)--(t4);
  \draw[edge] (t2)--(b1)--(b2)--(t3);

  \node[vertex, fill=#1] at (t1) {};
  \node[vertex, fill=#2] at (t2) {};
  \node[vertex, fill=#3] at (t3) {};
  \node[vertex, fill=#4] at (t4) {};
  \node[vertex, fill=#5] at (b1) {};
  \node[vertex, fill=#6] at (b2) {};
}

\begin{scope}
  \begin{scope}[shift={(0,0)}]
    \graphB{blue}{red}{blue}{red}{blue}{red}
  \end{scope}

  \begin{scope}[shift={(4,0)}]
    \graphB{red}{blue}{blue}{red}{blue}{red}
  \end{scope}

  \begin{scope}[shift={(0,-3)}]
    \graphB{red}{blue}{blue}{red}{red}{blue}
  \end{scope}

  \begin{scope}[shift={(4,-3)}]
    \graphB{red}{blue}{red}{blue}{red}{blue}
  \end{scope}

\end{scope}

\end{tikzpicture}
}

\caption{Nash equilibria for Graph A (a) and B (b) obtained through classical simulation. Blue ones are free riders, red dots are contributing agents.}
\label{fig:1}
\end{figure}
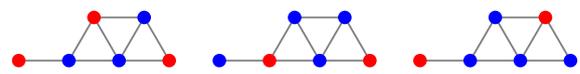
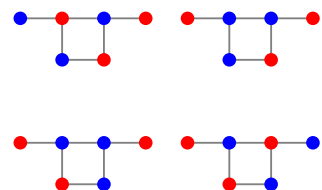

\section{Analysis and Nash Equilibrium search}
\label{sec:Analysis}
In light of what has been discussed in the prior sections, it is natural to propose interpreting the ground states (MIS) of an atomic 
Rydberg array -- characterized as a unit-disk graph due to the radius $R_b$ -- as Nash equilibria for a public goods game on that network. 
However, in this context, the network among agents must be reinterpreted and mapped such that agents are considered neighbors 
according to the geometric constraints of a unit-disk graph. This approach requires first mapping the network to a unit-disk graph -- a non-trivial task -- and then implementing this mapping
 while adhering to the physical constraints of a neutral atom machine. These challenges will be explored in the \textit{Mappable Network Games} section, while the \textit{Simulations} section will present comparisons between Nash equilibrium solutions and those obtained through bloqade simulations.

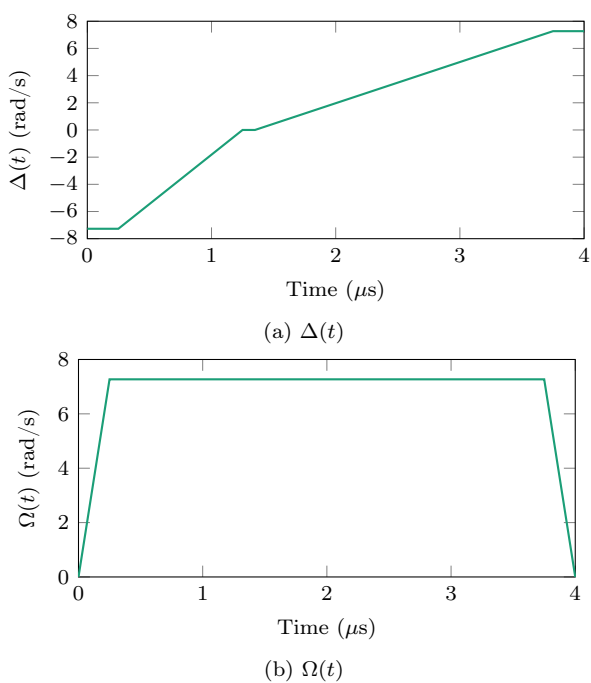
\begin{figure}[H]
  \centering
  \begingroup
    \setlength{\intextsep}{4pt}
    \setlength{\abovecaptionskip}{3pt}
    \setlength{\belowcaptionskip}{0pt}

    \subfloat[$\Delta(t)$]{%
      \begin{tikzpicture}
        \begin{axis}[
          width=\columnwidth,
          height=0.55\columnwidth,
          xmin=0, xmax=4,
          ymin=-8, ymax=8,
          xlabel={Time ($\mu$s)},
          ylabel={$\Delta(t)$ (rad/s)},
          xtick={0,1,2,3,4},
          ytick={-8,-6,-4,-2,0,2,4,6,8},
          tick label style={font=\footnotesize},
          label style={font=\footnotesize},
        ]
          \addplot[thick, bloqadegreen]
            coordinates {
              (0.00, -7.27)
              (0.25, -7.27)
              (1.25,  0.00)
              (1.35,  0.00)
              (3.75,  7.27)
              (4.00,  7.27)
            };
        \end{axis}
      \end{tikzpicture}
    }\\[-0.4ex]

    \subfloat[$\Omega(t)$]{%
      \begin{tikzpicture}
        \begin{axis}[
          width=\columnwidth,
          height=0.55\columnwidth,
          xmin=0, xmax=4,
          ymin=0, ymax=8,
          xlabel={Time ($\mu$s)},
          ylabel={$\Omega(t)$ (rad/s)},
          xtick={0,1,2,3,4},
          ytick={0,2,4,6,8},
          tick label style={font=\footnotesize},
          label style={font=\footnotesize},
        ]
          \addplot[thick, bloqadegreen]
            coordinates {
              (0.00, 0.00)
              (0.25, 7.27)
              (1.25, 7.27)
              (1.35, 7.27)
              (3.75, 7.27)
              (4.00, 0.00)
            };
        \end{axis}
      \end{tikzpicture}
    }

    \caption{Linear profiles used in Bloqade simulation}
    \label{fig:linear}
  \endgroup
\end{figure}


\subsection{Groundstates and Nash Equilibrium correspondence}
The reasoning behind this correspondence can be intuitively understood by observing how the Rydberg blockade mechanism and 
the strategic substitutes effect produce similar outcomes on a graph. Just as the Rydberg blockade triggered by node \(i\) prevents 
neighboring atoms from becoming excited, a node’s decision to contribute in the public goods game discourages its neighbors from doing the same. 
These are local interactions -- agents and atoms influencing only their neighbors -- with analogous effects that result in MIS configurations. It is worth noting that this correspondence relies on the assumption that the graph supporting the game is a unit-disk graph. 
While this assumption partially limits the generality of the claim, it provides a useful framework for understanding and formalizing the phenomenon.

\subsection{Simulations}
\begin{figure*}[ht]
    \centering

    \subfloat[][]{
        \includegraphics[width=0.4\textwidth]{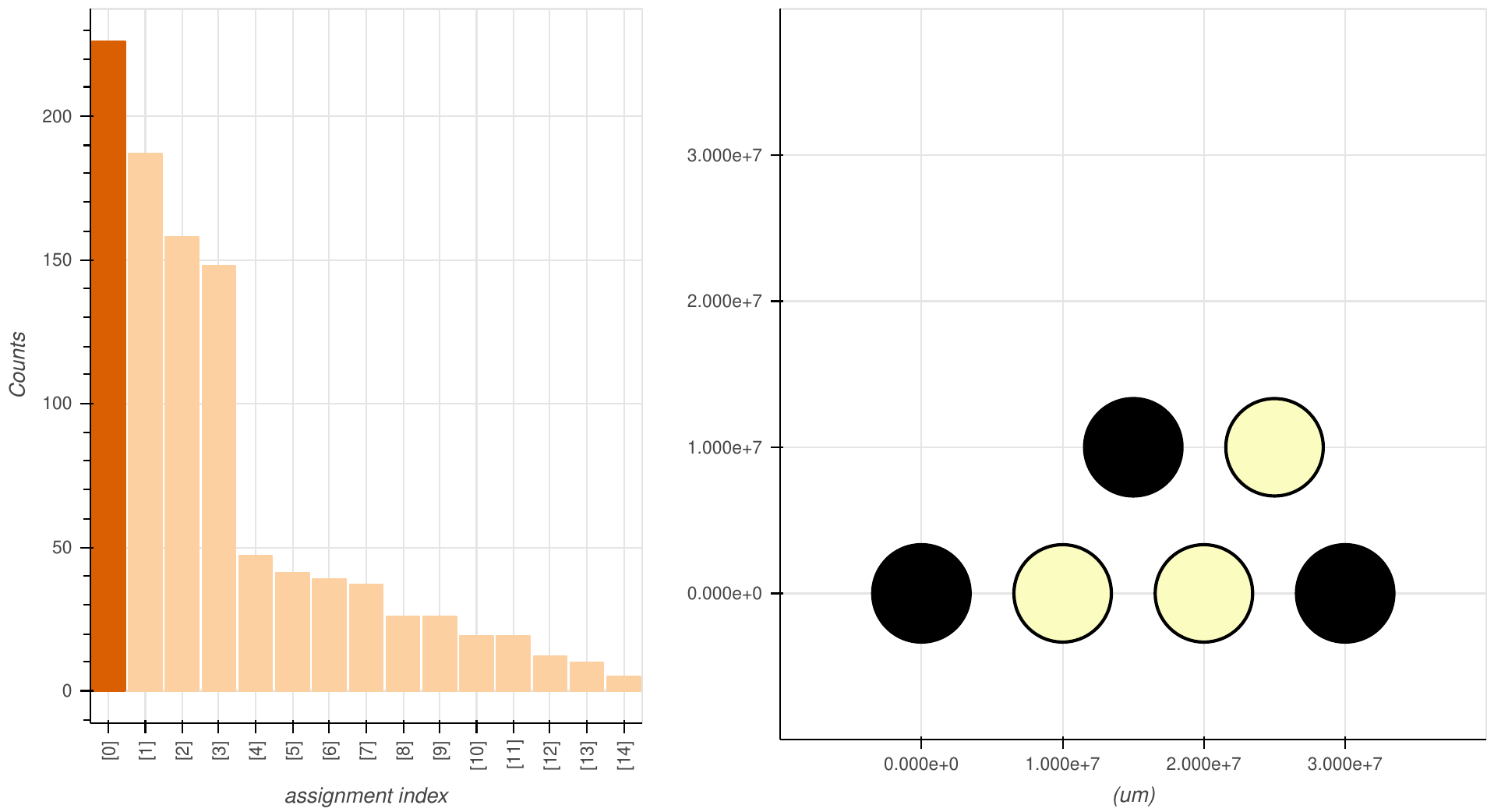}
        \label{fig:subimage3}}
    \subfloat[][]{
        \includegraphics[width=0.4\textwidth]{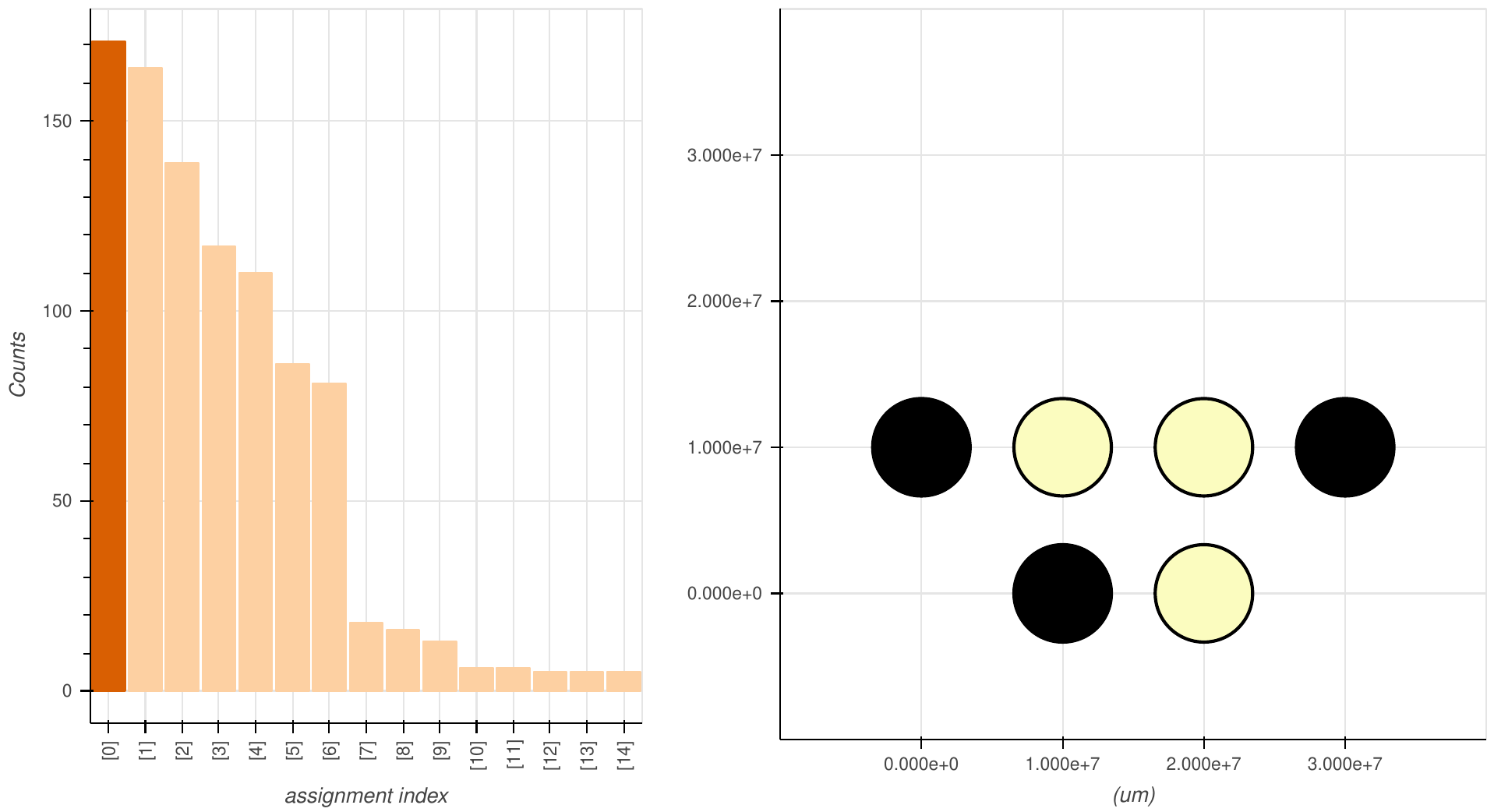} 
        \label{fig:subimage4}}\\
    \subfloat[][]{
        \includegraphics[width=0.4\textwidth]{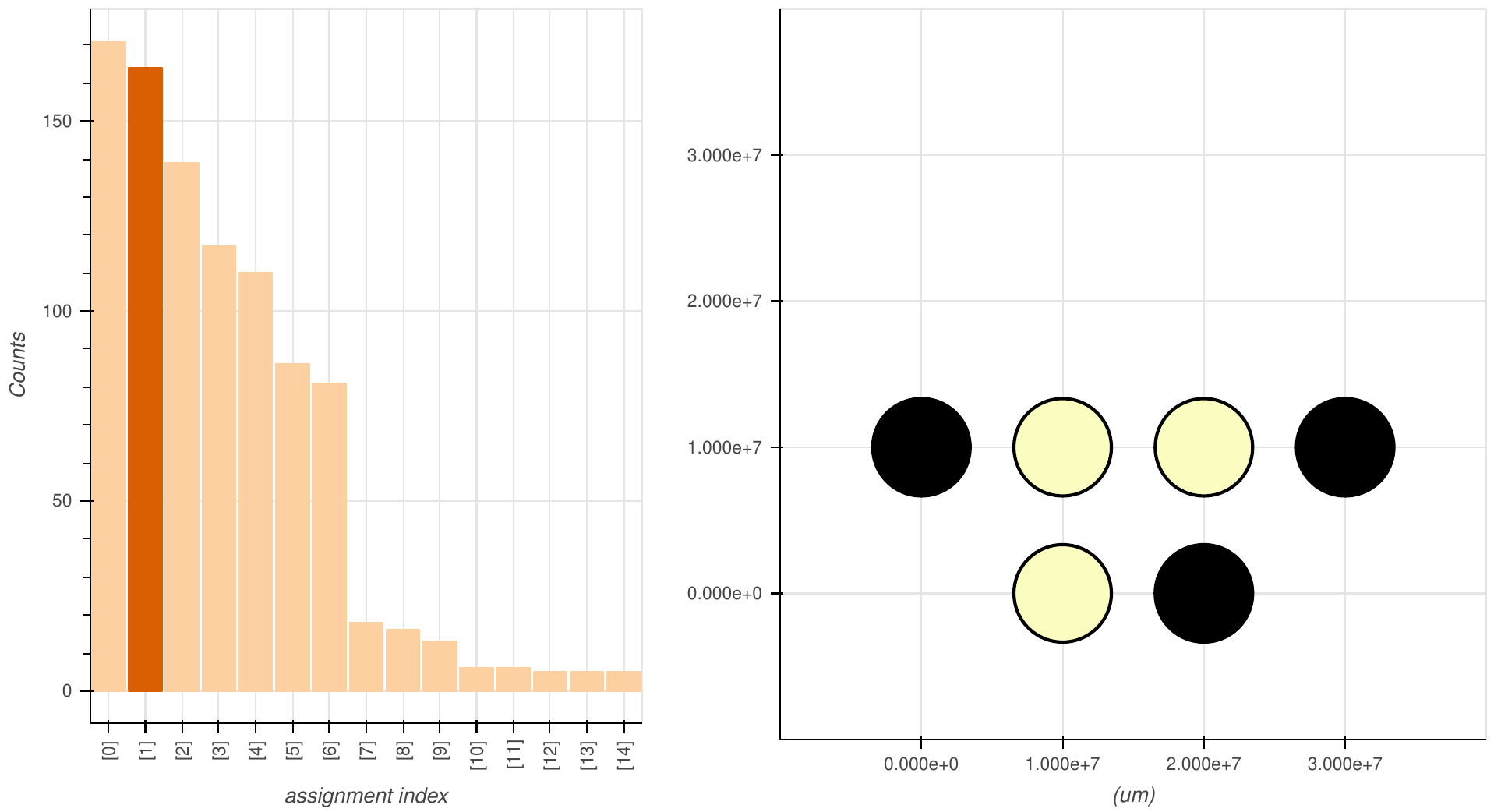} 
        \label{fig:subimage5}}\\
    \subfloat[][]{
        \includegraphics[width=0.4\textwidth]{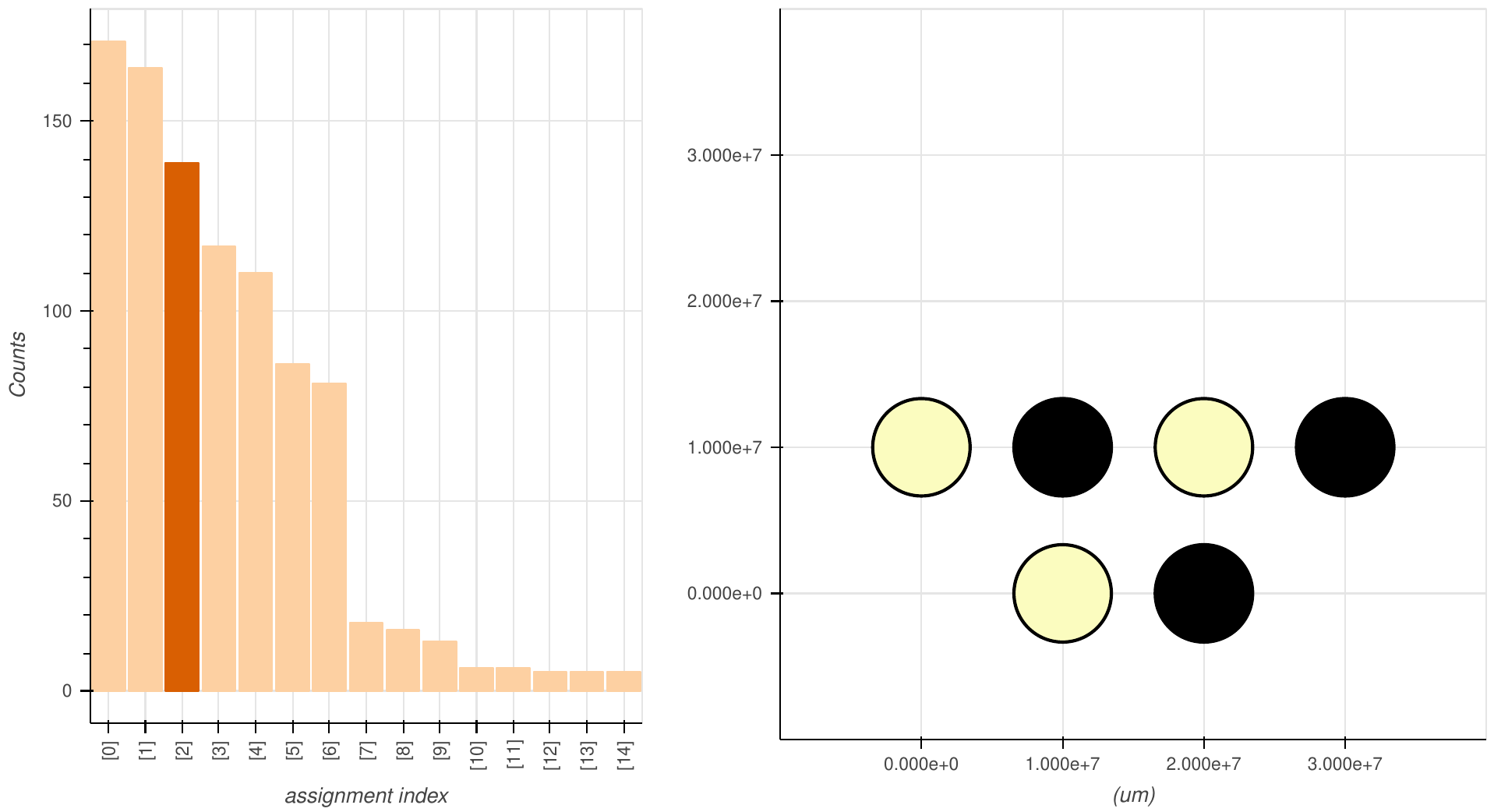} 
        \label{fig:subimage6}}
    \subfloat[][]{
        \includegraphics[width=0.4\textwidth]{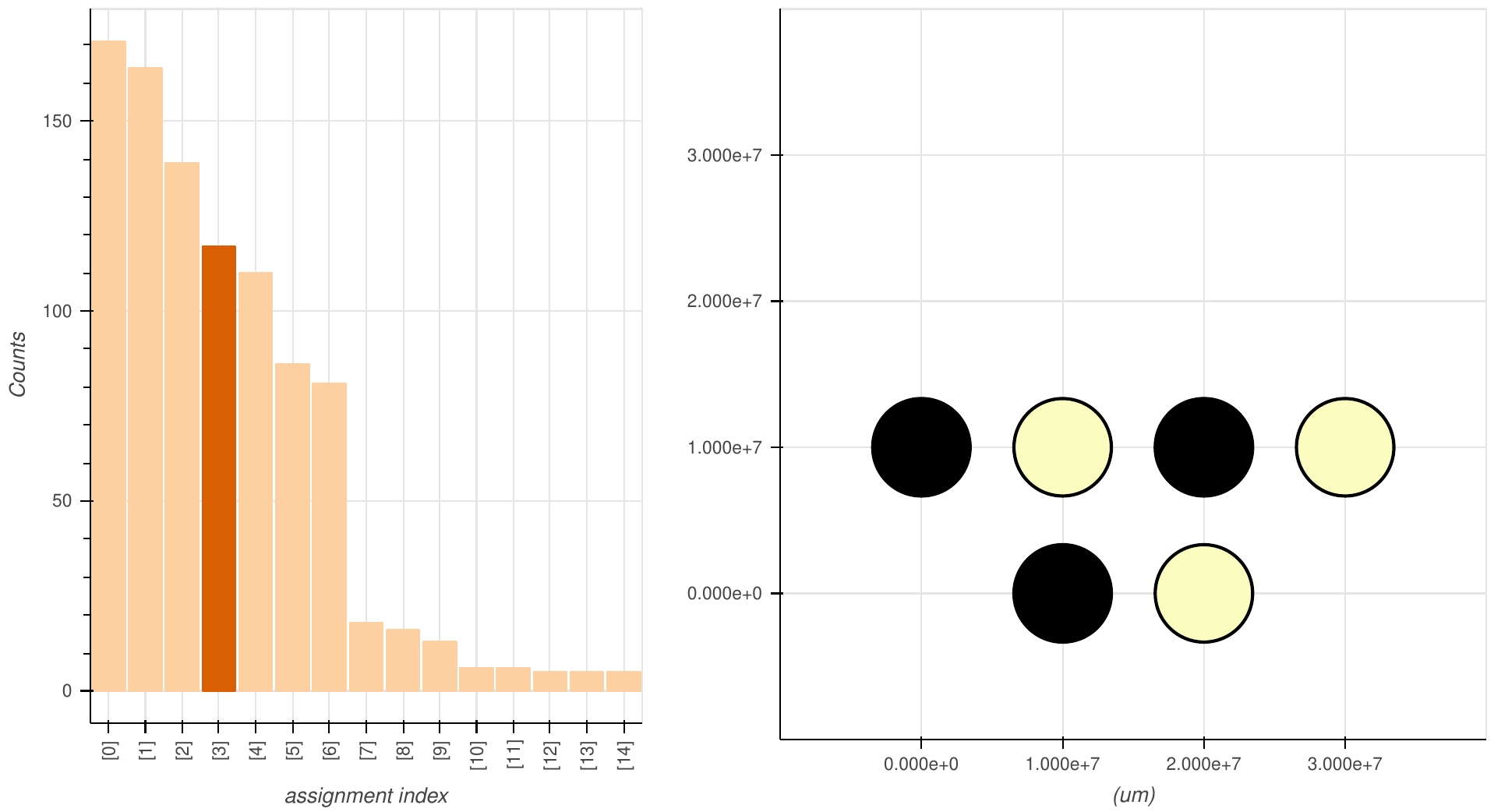} 
        \label{fig:subimage7}}

    \caption{Bloqade simulation for 1000 shots, zoomed in on the most frequently observed configurations of graph A (a) 
    and graphs B (b, c, d, e). The full line in the histogram corresponds to the configuration shown on the right.
     Black dots represent excited atoms, which form an MIS. All the displayed configurations are ground states of their respective systems.}
    \label{fig:simul}
\end{figure*}

For our simulations, we used the Bloqade library. This library is designed to construct and solve quantum systems that can be simulated on quantum hardware,
such as the Aquila machine (QUERA), which we took as a reference. We had to account for the various technological limitations of the machine,
such as the minimum distance between two atoms, the maximum duration of user-defined evolution, and the maximum values of \(\Omega\) and \(\Delta\) \cite{Aquila_Quera}. We chose to simulate two different graphs: one with a unique MIS (Graph A) Figure \ref{fig:subimage1}, and another with four MIS (Graph B) Figure \ref{fig:subimage2}. 
As shown in Figure \ref{fig:simul}, most of the simulation results fall within the left peak of the graph, corresponding to the desired configuration: the MIS of our system. In Figure \ref{fig:simul}, we plotted the results of a simulation with 1000 shots, where the histograms provide a statistical representation of the final states. As we can see from Graph A, the peak Figure \ref{fig:subimage3}, which we interpret as the ground state, presents a configuration where the excited atoms form a MIS. For Graph B, since it has multiple MIS, we should instead expect a degenerate ground state. Indeed, we can observe that the corresponding configurations are the top four in terms of counts, as shown in Figure \ref{fig:simul}. It is worth noting that the linear profiles were crucial for obtaining valid results in Figure \ref{fig:linear}. Ensuring that the evolution time remained below \SI{4}{\micro\second} \cite{Aquila_Quera}, small variations in the steepness of the slopes had noticeable or slight effects on the final state statistics. In general, the most frequently observed configuration corresponded to a MIS. While this was not an issue for Graph A, it posed a challenge for Graph B.
Specifically, the degenerate nature of Graph B's ground state was not always reflected in these configurations being the most frequently observed. This is likely due to the fact that, while the annealing schedule on Aquila remains consistent throughout the evolution, it is not flawless.
Diabatic transitions, noise, and decoherence effects can introduce imperfections. As a result, some measurements may violate the independent set condition or fail to correspond to maximal independent sets \cite{Aquila_Quera}. To complete our analysis, we simulated a public goods game and identified all Nash equilibria using the previously defined payoff function. For this, we generated all possible strategy combinations and evaluated each to determine whether it qualified as a Nash equilibrium. 
This involved comparing the payoff from the current strategy with that obtained by switching to an alternative strategy. 
This exhaustive approach ensures that all Nash equilibria are identified without omission. The results of these simulations, pictorially shown in Figure \ref{fig:subimage1}, reveal that Graph A has five Nash equilibria (mIS), with only one qualifying as a maximal independent set (MIS). The red dots represent the contributing agents, while the blue ones represent the free-riders. 
For Graph B, all Nash Equilibria are MIS (Figure \ref{fig:subimage2}). It's easy to see that the found configurations are the same for both simulations. An additional verification was conducted using a classical search algorithm to identify all the mISs on a unit disk graph. The algorithm generated all possible configurations to ensure both independence and maximality. The code used for our simulations is publicly available on GitHub

\subsection{Mappable Network Games}
\label{subsec:Map}
The problem of mapping is of primary importance and significantly restricts 
the available possibilities. Some of these constraints stem from 
machine technology, while others arise from the topology of the graph. 
Firstly, it is important to emphasize that "to map" (or "to embed") here means to 
create a geometric configuration of atoms such that each atom corresponds to a 
node of the graph being mapped, and every edge between two nodes is represented by the 
link between two atoms. Technologically, this can be achieved in the 
Aquila quantum computer using optical tweezers to manipulate the atoms \cite{Aquila_Quera}.
Since a two-dimensional lattice of neutral atoms is being utilized, where 
atoms can only be arranged in two-dimensional configurations, the mapping 
is restricted to two-dimensional graphs.
Another constraint arises from the limited freedom with which we can 
arrange atoms. Specifically, according to \cite{Aquila_Quera}, in the 
Quera 256-qubit machine, \textit{Aquila}, neither the row spacing nor 
the distance between two atoms in the same row can be less than \SI{4}{\micro\meter}. 
These constraints typically have a smaller impact on regular lattices but may 
become more significant when designing arbitrary topologies. Additionally, the row 
spacing limitation was introduced as a design choice to accelerate the 
sorting process, though it is not a fundamental restriction and could, 
in principle, be relaxed.
Another issue is related to the geometrical constraints involved in 
creating edges between atoms. As pointed out in \cite{Ellis2007}, in 
order to guaranty a connection between two nodes (i.e., atoms) within a 
certain distance, the graph must be at most linearly dense. 
Moreover, even if the $4\,\mu$m limit has been respected, connectivity 
issues can arise due to the extreme proximity of the two shells determined 
by the Rydberg radii $R_b$ \cite{Vercellino_et_al}.
In fact, if two atoms that are not intended to be connected have their "Rydberg 
shells" too near, i.e., their distance is $\eta \gtrsim 2R_b$, there is some 
probability that the two atoms will link due to noise. Another embedding problem 
arises from the connectivity procedure used by these atoms. It is important to 
remember that when increasing the detuning parameter $\Delta(t)$, the atoms begin to 
arrange themselves into the desired configuration by interacting due to the 
increasing Rydberg radius. For a large graph, the time for the atoms to settle may 
exceed the decoherence time, making the algorithm less efficient and precise 
\cite{Vercellino_et_al}.
It is important to note that the difference between the adjacency distance, defined 
as the maximum distance at which two atoms link, and the non-adjacency distance, 
which is the minimum distance between two "Rydberg shells" that prevents linking, 
varies with the number of graph nodes, i.e., the atoms used to represent the graph. 
As highlighted in \cite{Quantum_Opt_for_MIS}, the quantum algorithm must be 
specifically optimized for each graph. In particular, it is essential to examine the 
slowdown point of the detuning parameter $\Delta(t)$ in order to maximize the 
probability of identifying a MIS.

\subsection{Computational time}

Another important topic, arguably the most critical in this case, is the computational 
time.  As reported in \cite{Aquila_Quera}, the machine can perform all 
calculations in a single step, so we could assume the computational time is $O(1)$, as 
the steps performed are independent of the number of nodes $n$. However, 
this assumption is only approximately valid for graphs with a very limited number of nodes, 
since the execution time per step, i.e., the time it takes for the machine to 
perform a single operation, is in fact influenced by the number of nodes (i.e., neutral atom qubits). 
It has been discussed previously that a larger number of nodes 
implies a longer time for the atoms to settle \cite{Vercellino_et_al}, meaning the 
computational time increases 
until it reaches the decoherence time, at which point the algorithm becomes unusable. 
As pointed out in \cite{Quantum_Opt_for_MIS}, the quantum algorithm's performance is also 
highly (even if not exclusively) influenced by the minimum energy gap $\delta_{\text{en}}$, 
i.e., the smallest difference in energy between the quantum states involved in the algorithm. 
In fact, under certain conditions, it is possible, according to \cite{Quantum_Opt_for_MIS}, 
to achieve a quadratic speedup for certain values of $\delta_{\text{en}}$ with respect to the 
classical hardness parameter of the problem, despite the fact that this relationship does not 
always hold.

\section{Conclusions}
\label{sec:conclusions}

Neutral atom quantum computers present a promising platform for studying complex network dynamics, such as Public Goods on Networks systems. Their unique physical mechanisms make them particularly well-suited for exploring the intersection of game theory and quantum computation. On a unit disk graph, both the game and the physical system follow analogous rules, resulting in outcomes that manifest as Maximum Independent Sets (MIS). A key aspect of this approach lies in the mapping process, where the network structure of the game is embedded into the physical layout of the atomic system. This requires careful tuning and the resolution of technical limitations to ensure accurate mapping and effective evolution toward the desired MIS configuration. It leverages the distinct capabilities of neutral atom quantum processors, which are particularly well-suited for tackling complex optimization problems. By explicitly mapping the problem of finding Nash equilibria onto a neutral atom quantum computing framework, this work introduces a novel method that has not been extensively examined in the existing literature. This research advances earlier work by addressing the computational challenges of finding Nash equilibria through a relatively new quantum computing framework. Unlike classical or heuristic methods, this study demonstrates how quantum systems can directly solve these problems, offering a more efficient and scalable approach.
\bibliographystyle{ieeetr}
{\small \bibliography{Referencies}}
\end{multicols}
\end{document}